\documentclass[onecolumn,reviewcopy]{autart}

\usepackage{amssymb}
\usepackage{amsmath}
\usepackage{natbib}
\usepackage{color}
\usepackage{graphics}
\usepackage{graphicx}
\usepackage[tight,footnotesize]{subfigure}
\usepackage{multirow}
\usepackage{algorithm,algorithmic}
\usepackage{savesym}
\savesymbol{AND}

\usepackage{tikz}
\usetikzlibrary{fit,shapes.misc}

\def \L{\mathcal{L}}
\def \E{\mathcal{E}}
\def \V{\mathcal{V}}

\def \S{\mathcal{S}}

\def \D{\mathcal{D}}
\def \A{\mathcal{A}}

\def \C{\mathcal{C}}

\def \P{\mathcal{P}}

\def \M{\mathcal{M}}

\def \Z{\mathbb{Z}}

\newtheorem{remark}{Remark}
\newtheorem{lemma}{Lemma}
\newtheorem{definition}{Definition}
\newtheorem{corollary}{Corollary}

\begin{document}

\begin{frontmatter}

\title{Speeding up finite-time consensus via minimal polynomial of a weighted graph - a numerical approach}

\author[AddA]{Zheming Wang} \ead{wang.zheming@nus.edu.sg},
\author[AddA]{Chong Jin Ong}\ead{mpeongcj@nus.edu.sg}

\address[AddA]{Department of Mechanical Engineering, National University of Singapore, 117576, Singapore}


\begin{abstract}
Reaching consensus among states of a multi-agent system is a key requirement for many distributed control/optimization problems.
Such a consensus is often achieved using the standard Laplacian matrix (for continuous system) or Perron matrix
(for discrete-time system). Recent interest in speeding up consensus sees the development of finite-time consensus algorithms.
This work proposes an approach to speed up finite-time consensus algorithm using the weights of a weighted Laplacian matrix.
The approach is an iterative procedure that finds a low-order minimal polynomial that is consistent with the topology of the underlying graph. In general, the lowest-order minimal polynomial achievable for a network system is an open research problem. This work proposes a numerical approach that searches for the lowest order minimal polynomial via a rank minimization problem using a two-step approach: the first being an optimization problem involving the nuclear norm and the second a correction step. Several examples are provided to illustrate the effectiveness of the approach.
\end{abstract}

\end{frontmatter}




%






\section{Introduction}

Achieving consensus of states is a well-known important feature for networked system, see for example \citep{ART:OM04,BOO:RB07}. Many distributed control/optimization problems over a network require a consensus algorithm as a key component.
The most common consensus algorithm is the dynamical system defined by the Laplacian matrix for continuous time system or the Perron matrix for discrete-time system. Past works in the general direction of speeding up convergence of these algorithms exist.  For example, the work of \citep{ART:XB04} proposes a semi-definite programming approach to minimize the algebraic connectivity over the family of symmetric matrices that are consistent with the topology of the network. Their approach, however, results in asymptotic convergence towards the consensus value and is most suitable for larger networks. More recent works focus on finite-time convergence consensus algorithm  \citep{INP:SH07,INP:YSSG09, ART:WX10,ART:YSSBG13,ART:HJOV14,ART:HSJ15} which is generally preferred for small to moderate size networks. One important area in finite-time convergence literature is the determination of the asymptotic value of a consensus network using a finite number of state measurement. Typically, the approach adopted is based on the z-transform final-value theorem and on the finite-time convergence for  individual node \citep{INP:SH07,INP:YSSG09,ART:YSSBG13}.  Other works in finite-time consensus include the design of a short sequence of stochastic matrices $A_k, \cdots, A_0$ such that $x(k)= \Pi_{j=1}^{k} A_j x(0)$ reaches consensus after $k$ steps \citep{INP:KS09,ART:HSJ15}.

This work proposes an approach to speed up finite-time convergence consensus algorithm for a network of agents via the weights associated with the edges of the graph. It is an offline method where the network is assumed known. Hence, it is similar in spirit to the work of \cite{ART:XB04} except that the intention is to find a low-order minimal polynomial. Ideally, the lowest-order minimal polynomial should be used. However, the lowest minimal polynomial achievable for a given graph with variable weights is an open research problem \citep{ART:FH07}. They are only known for some special classes of graphs (full connected, star-shaped, strongly regular and others), \citep{ART:VH98,ART:VKT14}. For this reason, this paper adopts a computational approach towards finding a low-order minimal polynomial. The proposed approach achieves the lowest order minimal polynomial in many of the special classes of graphs and almost always yields minimal polynomial of order lower than those obtained from standard Perron matrices of general graphs. These are demonstrated by several numerical examples.

The choice of the weights to be chosen is obtained via a rank minimization problem. In general, rank minimization is a well-known difficult problem \citep{INP:FHB04,ART:RFP10}. Various approaches have been proposed in the literature including the nuclear norm relaxation, bilinear projection methods and others. This work uses a unique two-step procedure in the rank minimization problem. The first is an optimization problem using the nuclear norm and the second, which uses the results of the first, is a correction step based on a low rank approximation.
While both steps of the two-step procedure have appeared in the literature, the use of the two in a two-step procedure is novel, to the best of the authors' knowledge. Hence, the approach towards the rank minimization problem can be of independent interest. The same can be said of the expression of the consensus value which is obtained not by the z-transform mechanization.

The rest of this paper is organized as follows. This section ends with a description of the notations used. Section \ref{sec:pre} reviews some standard results of the standard Laplacian and Perron matrix with unity edge weights as well as minimal polynomial and its properties. Section \ref{sec:main} presents the procedure of obtaining the consensus value from the minimal polynoimal and discusses, in detail, the key subalgorithm used in the overall algorithm. The overall algorithm is described in section \ref{sec:algo} and the performance of the approach is illustrated via several numerical examples in Section \ref{sec:num}. Conclusions is given in Section \ref{sec:con}.

The notations used in this paper are standard. Non-negative and positive integer sets are indicated by $\mathbb{Z}^+_0$ and $\mathbb{Z}^+$ respectively. Let $M, L \in \Z^+_0$ with $M \ge L \ge 1$. Then $\mathbb{Z}^M:=\{1,2,\cdots,M\}$ and $\mathbb{Z}_L^M:=\{L,L+1,\cdots,M\}$. Similarly, $\mathbb{R}^+_0$ and $\mathbb{R}^+$ refer respectively to the sets of non-negative and positive real number. $I_n$ is the $n\times n$ identity matrix, $1_n$ is the $n$-column vector of all ones (subscript omitted when the dimension is clear). Given a set $C$, $|C|$ denotes its cardinality.  The transpose of matrix $M$ and vector $v$ are indicated by $M'$ and $v'$ respectively. For a square matrix $Q$, $Q \succ (\succeq) 0$ means $Q$ is positive definite (semi-definite), $spec(Q)$ refers to the set of its eigenvalues, $vec(Q)$ is the representation of elements of $Q$ as a vector and $(\lambda, v)$ is an eigen-pair of $Q$ if $Qv=\lambda v$. The cones of symmetric, symmetric positive semi-definite and symmetric and positive definite matrices are denoted by $\S^n=\{M \in \mathbb{R}^{n\times n} | M=M'\}$, $\S^n_{0+}=\{M \in \mathbb{R}^{n\times n} | M=M', M \succeq 0\}$ and $\S^n_+=\{M \in \mathbb{R}^{n\times n} | M=M', M \succ 0\}$ respectively.  The $\ell_p$-norm of $x\in \mathbb{R}^{n}$ is $\|x\|_p$ for $p=1,2,\infty$ while $\|M\|_*, \|M\|_2, \|M\|_F$ are the nuclear, operator (induced) and Frobenius norm of matrix $M$. Diagonal matrix is denoted as $diag\{d_1,\cdots,d_n\}$ with diagonal elements $d_i$. Additional notations are introduced when required in the text.

\section{Preliminaries and Problem formulation}\label{sec:pre}
This section begins with a review of standard consensus algorithm and sets up the notations needed for the sequel.
The network of $n$ nodes is described by an undirected graph $G = (\V,\E)$ with vertex set $\V = \{1,2,\cdots,n\}$ and edge set $\E \subseteq \V \times \V$. The pair $(i,j)\in \E$ if $i$ is a neighbor of $j$ and vice versa since $G$ is undirected. The set of neighbors of node $i$ is $N_i:=\{j\in \V: (i,j) \in \E, i \neq j\}$. The standard adjacency matrix $\A_s$ of $G$ is the $n \times n$ matrix whose $(i,j)$ entry is $1$ if $(i,j)\in \E$, and $0$ otherwise. 

The implementation of the proposed consensus algorithm is a discrete-time system of the form $z(k+1)=\P z(k)$ where $\P$ is the Perron matrix. However, for computational expediency, the working algorithm uses the weighted Laplacian matrix $\L \in \S^n_{0+}$. The conversion of $\L$ to $\P$ is standard and is discussed later, together with desirable properties of $\P$ and $\L$.
The properties of standard (non-weighted) $\L$ are first reviewed.

The standard Laplacian matrix denote by $\L_s$ of a given $G$ is
\begin{align} \label{eqn:Lii}
[\L_s]_{i,j}=\left\{
  \begin{array}{ll}
    -1, & \hbox{if $j \in N_i$;} \\
    |N_i|, & \hbox{if $i=j$ ;} \\
    0, & \hbox{otherwise.}
  \end{array}
\right.
\end{align}
In this form, it is easy to verify that (i) eigenvalues of $\L_s$ are real and non-negative, (ii) eigenvectors corresponding to different eigenvalues are orthogonal, (iii) $\L_s$ has at least one eigenvalue $0$ with eigenvector of $1_n$. Properties (i) and (ii) follow from the fact that $\L_s$ is symmetric and positive semi-definite while properties (iii) is a result of the row sum of all rows being $0$.
Suppose the assumption of \\
$(A1)$: $G$ is connected. \\
is made. Then, it is easy to show that the eigenvalue of $0$ is simple with eigenvector $1_n$. Consequently, the consensus algorithm of
$\dot{x}(t)=-\L_s x(t)$ converges to $\frac{1}{n} 1_n (1_n' x(0))$.

Unlike standard Laplacian, this work uses the weighted Laplacian
\begin{align}
\L(W,G)= \D(G)- \A(G,W)
\end{align}
where $\A(G,W)$ is the weighted adjacency matrix with $[\A(G,W)]_{ij}=w_{ij}$ when $(i,j) \in \E$,
$\D(G)=diag\{d_1,d_2,\cdots, d_n\}$ with $d_i:=\sum_{j\in N_i} w_{ij}$ and $W:=\{w_{ij} \in \mathbb{R} | (i,j) \in \E\}$.
The intention of this work is to compute algorithmically the minimal polynomial of $\L(W,G)$ over variable $W$ for a given $G$.
However, since the minimal polynomial attainable for a given network $G$ is a well-known difficult problem \citep{ART:FH07},
the output of the algorithm can be seen as an upper bound on the order of the achievable minimal polynomials of $\L(W,G)$ over all $W$.
Note that the value of $w_{ij}$ is arbitrary including the possibility that $w_{ij}=0$ and $w_{ij}<0$ for $(i,j) \in \E$.
This relaxation allows for a larger $W$ search space but it brings about the possibility of losing connectedness of $\L(W,G)$ even when $G$ is connected. Additional conditions are therefore needed to preserve connectedness, as discussed in the sequel. Since the graph $G$ is fixed, its dependency in $\L(\cdot), \D(\cdot)$ and $\A(\cdot)$ is dropped for notational convenience hereafter unless required.

The desirable properties of $\L(W)$ are
\begin{itemize}
  \item[(L1)] All eigenvalues are non-negative.
  \item[(L2)] 0 is a simple eigenvalue with eigenvector $1_n$.
  \item[(L3)] $[\L(W)]_{ij}=0$ when $(i,j) \notin \E$
\end{itemize}
Properties (L1) and (L2) are needed for $x(t)$ of $\dot x(t)=-\L x(t)$ to reach consensus while property (L3) is a hard constraint imposed by the structure of $G$. In addition, the choice of $W$ should be chosen such that
\begin{itemize}
  \item[(L4)] $\L(W)$ has a low-order minimal polynomial.
\end{itemize}

Given $\L(W)$ having properties (L1)-(L4), the corresponding Perron matrix $\P$ can be obtained from $\P:=e^{-\epsilon\L}$ or
$\P :=I_n - \epsilon \L(W)$, for some $0<\epsilon < \frac{1}{max_i \{d_{i}\}}$). Then, it is easy to verify that $\P$ inherits
from (L1)-(L4) the following properties:
\begin{description}
  \item (P1) All eigenvalues of $\P$ are real and lie within the interval $(-1,1]$.
  \item (P2) $1$ is a simple eigenvalue of $\P$ with eigenvector $1_n$.
  \item (P3) $[\P]_{ij}=0$ when $(i,j) \notin \E$.
  \item (P4) $\P$ has a low-order minimal polynomial.
\end{description}
The consensus algorithm based on $\P$ follows
\begin{align}
z(k+1)=\P z(k) \label{eqn:zkplus1}
\end{align}
for consensus variable $z \in \mathbb{R}^n$. From (P1)-(P2) and assumption (A1), it is easy to show, with $(\sigma_i,\xi_i)$ being the $i^{th}$ eigenpair of $\P$ that
\begin{align}
lim_{k \rightarrow \infty} z(k) = lim_{k \rightarrow \infty} (\sum_{i=1}^n \xi_i \xi_i' \sigma_i^k)z(0) =\frac{1}{n} 1_n (1_n' z(0))
\end{align}
and hence $lim_{k \rightarrow \infty} z(k)$ reaches consensus among all its elements. The above shows that finding a $\P$ that possesses properties (P1)-(P4) is equivalent to finding an $\L(W)$ that satisfying properties (L1)-(L4). The next subsection reviews properties of the minimal and characteristic polynomials that are available in the literature.

\subsection{Minimal polynomial and finite-time convergence}
This section begins with a review of minimal polynomial and its properties.
\begin{definition}
The minimal polynomial $m_M(t)$ of a square matrix $M$ is the monic polynomial of the lowest order such that $m_M(M)=0$.
\end{definition}
Several well known properties of characteristic and minimal polynomial are now collected in the next lemma.
Their proofs can be found in standard textbook (see for example, \cite{BOO:FIS03} or others) and are given next for easy reference.
\begin{lemma} \label{lem:poly}
Given a square matrix $M$ with minimal polynomial $m_M(t)$ and characteristic polynomial $p_M(t)$. Then (i) $\lambda$ is an eigenvalue of $M$ if and only if $\lambda$ is a root of $m_M(t)$. (ii) a root of $m_M(t)$ is a root of $p_M(t)$ (iii) the distinct roots of $m_M(t)$ are equivalent to the distinct roots of $p_M(t)$.  Suppose $M$ is symmetric, then (iv) the algebraic multiplicity of every eigenvalue in $M$ equals its geometric multiplicity and (v) each distinct root of $m_M(t)$ has a multiplicity of one.
\end{lemma}

Since $G$ is undirected, the desirable properties of (L4) (or equivalently (P4)) is achieved by having as few distinct roots of $m_\L(t)$ as possible, as given in properties (iv) and (v) of Lemma \ref{lem:poly}.  The next result shows the connection of the minimal polynomial and the computations needed to obtain the consensus value of (\ref{eqn:zkplus1}) using only $s_\P$ number of states where $s_\P$ is the order of the minimal polynomial, $m_\P(t)$.

\section{Main approach} \label{sec:main}
\begin{lemma} \label{eqn:mpoly}
Given a $n^{th}$ order symmetric matrix $\P$ with minimal polynomial $m_{\P}(t)$ of order $s_\P$. Any matrix polynomial $h(\P)$ can be expressed as $h(\P)=r(\P)$ where $r(t)$ is a $(s_\P -1)$ order polynomial. The coefficients of $r(t)$ can be obtained from solution of
the following equation involving the Vandermonde matrix.
\begin{align} \label{eqn:ht}
\left(
  \begin{array}{ccccc}
    1 & \lambda_1 & \lambda_1^2 & \cdots & \lambda_1^{{s_\P}-1} \\
    \vdots &  & \ddots &  & \vdots \\
    1 & \lambda_{s_\P} & \lambda_{s_\P}^2 & \cdots & \lambda_{s_\P}^{{s_\P} -1} \\
  \end{array}
  \right)  \left(
             \begin{array}{c}
               \pi_0 \\
               \vdots \\
               \pi_{{s_\P}-1} \\
             \end{array}
           \right)
  =\left(       \begin{array}{c}
                  h(\lambda_1) \\
                  \vdots \\
                  h(\lambda_{{s_\P}-1}) \\
                \end{array}
              \right)
\end{align}
where $\lambda_i$ are the distinct eigenvalues of $\P$ and $\pi_i$ is the coefficient of $t^i$ in $r(t)$.
\end{lemma}
\textit{Proof :}
The polynomial $h(t)$ can be expressed, via long division by $m_{\P}(t)$, as
$$h(t)=\phi(t)m_{\P}(t)+r(t)$$
where $r(t)$ is the remainder of the division and, hence, is of order $s_\P-1$ (including the possibility that some coefficients are zero). Since the above holds for all values of $t$, it holds particularly when $t=\lambda_i$, the $i^{th}$ distinct eigenvalue of $\P$. This fact, together with $m_{\P}(t)|_{t=\lambda_i}=0$ from property (P1) of Lemma (\ref{lem:poly}), establishes each row of (\ref{eqn:ht}).
From property (P3) of Lemma \ref{lem:poly}, there are $s_\P$ distinct eigenvalues of $\P$ and there are $s_\P$ unknown coefficients in $r(t)$ whose values can be obtained from solving (\ref{eqn:ht}). To show that (\ref{eqn:ht}) always admits a solution, note that the determinant of the Vandermonde matrix is $\Pi_{1 \le i \le j \le {s_\P}-1} (\lambda_j - \lambda_i)$ and is non-zero since $\lambda_i$ are distinct. The above equation also holds for its corresponding matrix polynomial or $h(\P)=\phi(\P)m_{\P}(\P)+r(\P)$. That $h(\P)=r(\P)$ follows because $m_{\P}(\P)=0$.
$\Box$

Consider a given polynomial of $h(t)=\lim_{k \rightarrow \infty} t^{k}:=t^{\infty}$ and that $\P$ satisfies properties (P1)-(P4).
Lemma \ref{eqn:mpoly} states that there exists a polynomial $r(t)$ or order $s_\P-1$ such that
$t^{\infty}=\pi_{s_\P-1}t^{s_\P-1}+\cdots + \pi_{1}t+\pi_0$ and the values of $\{\pi_{s_\P-1},\cdots, \pi_0\}$ can be obtained from the $s_\P$ equations of (\ref{eqn:ht}). The right hand side of (\ref{eqn:ht}) has the properties that $h(t)|_{t=\lambda_i}=\lambda_i^{\infty}=0$
for all but one eigenvalues of $\P$ since $|\lambda_i|<1$ (property P1). The remaining eigenvalue of $\P$ is $\lambda_i=1$ and it yields $h(t)|_{t=\lambda_i}=\lambda_i^{\infty}=1$, following property (P2). Hence,
\begin{align} \label{eqn:W}
\P^{\infty}=\pi_{s_\P-1}\P^{s_\P-1}+\cdots + \pi_{1}\P+\pi_0 I_n.
\end{align}
Rewriting (\ref{eqn:zkplus1}) as $z(k)=\P^{k}z(0)$, one gets
\begin{align}
\lim_{k \rightarrow \infty} z(k) &=  \P^{\infty}z(0)=( \sum\limits_{\ell=0}^{s_\P-1} \pi_{\ell} \P^{\ell})z(0)= \sum\limits_{\ell=0}^{s_\P-1} \pi_{\ell}z(\ell) \label{eqn:sumtoT}
\end{align}
The application of Lemma \ref{eqn:mpoly} for distributed consensus algorithm is now obvious. Each agent $i$ stores the parameters $\{\pi_{s_\P-1},\cdots, \pi_0\}$ in its memory as well as $\{z_i(0), \cdots, z_i(s_\P-1)\}$, obtained from
\begin{align}
z_i(k+1)&=\sum_{j \in N_i} \P_{ij} z_j(k), \;  \textrm{ for } k=0,\cdots, s_\P-2. \nonumber
\end{align}
At the end of $k=s_\P -2$, agent $i$ takes the sum of $\pi_0 z_i(0) + \cdots + \pi_{s_{\P}-1} z(s_{\P}-1)$  which yields $\lim_{k \rightarrow \infty} z_i(k)$.

\subsection{Key Subalgorithms}
With the discussion above, it is clear that the choice of $\L$ (or $\P)$ should be chosen such that $m_{\L}(t)$ is the lowest-order minimal polynomial consistent with the network.  The numerical approach proposed here attempts to find the lowest order minimal polynomial by minimizing the number of distinct eigenvalues of $\L(W,G)$ over variable $W$, since the order of the minimal polynomial $m_{\L}(t)$ equals to the number of distinct eigenvalues of $\L$ from Lemma \ref{lem:poly}.

The overall scheme of the proposed algorithm is now described in loose terms for easier appreciation.
The numerical algorithm is iterative and at each iteration $k$, two subalgorithms are invoked producing two possible choices of $W$. The $W$ that results in $\L(W,G)$ having a lower order of minimal polynomial is then chosen as $\L(W_k,G)$. The two subalgorithms, OPA and OPB, are very similar in structure but serve different purposes: OPA searches for a new eigenvalue of $\L(W_k,G)$ with multiplicity of 2 or higher while OPB searches for additional multiplicity of eigenvalues that are already present in $\L(W_k,G)$. To accomplish this, two sets are needed: $\C_k$ containing the zero simple eigenvalue and distinct eigenvalues with multiplicities of 2 or higher in $\L(W_k,G)$ and $\M_k$ containing the multiplicities of the eigenvalues in $\C_k$. The set $\C_k$ has the property that $\lambda \in \C_{k}$ implies $ \lambda \in \C_{k+1}$. The remaining `free' eigenvalues in $\L(W_k,G)$ are then optimized again in the next iteration. An additional index function $\xi(\cdot):\mathbb{Z} \rightarrow \mathbb{Z}$ is needed to keep track of the cardinality of $\C_{k}$ in the sense that $\xi(k)=|\C_{k}|$. The overall scheme proceeds with decreasing order of the minimal polynomial and stops when no further repeated eigenvalue can be found. At each iteration, properties (L1) - (L3) and assumption (A1) are preserved from $\L(W_{k-1},G)$. The key steps at iteration $k$ are now discussed. For notational simplicity, $\L(W_k,G)$ is denoted as $\L_k$.

Iteration $k$ requires the following data as input from iteration $k-1$: the matrix $\L_{k-1}$; index function $\xi(k-1)$; the set $\C_{k-1}=\{\lambda_1, \lambda_2, ..., \lambda_{\xi(k-1)}\}$ with $\lambda_1=0$ and $\lambda_2,\cdots, \lambda_{\xi(k-1)}$ being distinct eigenvalues of $\L_{k-1}$; and the set of multiplicities  $\M_{k-1}:=\{m_1, m_2, \cdots, m_{\xi(k-1)}\}$ where $m_1=1$ and $m_i \ge 2$ is the multiplicity of $\lambda_i$ in  $\C_{k-1}$. Let
\begin{align}\label{eqn:qkminus1}
q_{k-1}:=\sum_{i=1}^{\xi(k-1)} m_i,\quad \bar{q}_{k-1}=n-q_{k-1}
\end{align}
corresponding to the number of fixed and free eigenvalues in $\L_{k-1}$.

Iteration $k$ starts by computing the eigen-decomposition of $\L_{k-1}$ in the form of $\L_{k-1}=Q_{k-1} \Lambda_{k-1} Q_{k-1}'$ where $\Lambda_{k-1}=\left(
                \begin{array}{cc}
                  D_c & 0 \\
                  0 & D_o \\
                \end{array}
              \right)$
with $D_c \in \S^{q_{k-1}}_{0+}$ being a diagonal matrix with elements, in the same order, as the eigenvalues in $\C_{k-1}$ including multiplicities,
and $D_o \in \S^{\bar{q}_{k-1}}_{+}$ being the diagonal matrix containing the remaining eigenvalues of $\L_{k-1}$.  Correspondingly, the $Q_{k-1}$ can be expressed as $[Q_{k-1}^c \: Q_{k-1}^o]$ of appropriate dimensions such that
\begin{align}\label{eqn:Lkminus1}
\L_{k-1}=\left(
           \begin{array}{cc}
             Q_{k-1}^c & Q_{k-1}^o \\
           \end{array}
         \right)\left(
                \begin{array}{cc}
                  D_c & 0 \\
                  0 & D_o \\
                \end{array}
              \right)\left(
                       \begin{array}{c}
                         (Q_{k-1}^c)' \\
                         (Q_{k-1}^o)' \\
                       \end{array}
                     \right)
\end{align}
Consider the parameterization of $\L$ by a symmetric matrix $M \in \S^{\bar{q}_{k-1}}_{+}$ in the form of
\begin{align} \label{eqn:HM}
H(M):= \left(
           \begin{array}{cc}
             Q_{k-1}^c & Q_{k-1}^o \\
           \end{array}
         \right) \left(\begin{array}{cc}
                  D_c & 0 \\
                  0 & M \\
                 \end{array}
                \right) \left(
                       \begin{array}{c}
                         (Q_{k-1}^c)' \\
                         (Q_{k-1}^o)' \\
                       \end{array}
                     \right)=Q_{k-1}^c D_c (Q_{k-1}^c)' + Q_{k-1}^o M(Q_{k-1}^o)'
\end{align}
The structural constraints of the graph $G$ are imposed on $M$ via $[H(M)]_{ij}=[Q_{k-1}^c D_c (Q_{k-1}^c)']_{ij}
+ [Q_{k-1}^o M(Q_{k-1}^o)']_{ij}=0$ for $(i,j) \notin \E$. The collection of these structural constraints can be stated as $\Phi_{k-1} vec(M) = b_{k-1}$ where $vec(M)$ is the vectorial representation of $M$ with $\Phi_{k-1}$ and $b_{k-1}$ being the collection of appropriate terms from $Q_{k-1}^c D_c (Q_{k-1}^c)'$ and  $Q_{k-1}^o M(Q_{k-1}^o)'$ respectively. Consider the following optimization problems over variables
$\lambda \in \mathbb{R}, M \in \S^{\bar{q}_{k-1}}_{+}$:
\begin{subequations}\label{eqn:M}
\begin{align}
(OP) \quad \quad min  \quad &rank(\lambda I - M) \label{eqn:minrank}\\
&M \succ 0 \label{eqn:Mpositive}\\
&\Phi_{k-1} vec(M) = b_{k-1} \label{eqn:Phi}
\end{align}
\end{subequations}
Then, the next lemma summarizes its properties.
\begin{lemma} \label{lem:prop1}
Suppose $\L_{k-1}$ satisfies (L1)-(L3) and (A1). Then
(i) OP has a feasible solution. (ii) $spec(H(M^*))=spec(M) \cup spec(D_c)$. (iii) Suppose $(\lambda^*, M^*)$ is the optimizer of OP. Then $H(M^*)$ satisfies (L1)-(L3) and (A1).

\end{lemma}
\textit{Proof:} (i) Choose $\hat{M}=D_o$ where $D_o$ is that given in (\ref{eqn:Lkminus1}) and $\hat{\lambda}$ be any diagonal element of $D_o$. Then $(\hat{\lambda},\hat{M})$ is a feasible solution to OP since $\L_{k-1}$ satisfies (L1)-(L3) and (A1). (ii) The property is obvious from the expression of $H(M^*)$ of (\ref{eqn:HM}). (iii) From (ii), the spectrum of $H(M^*)$ are the eigenvalues in $\C_{k-1}$ (with the corresponding multiplicities) and those of $M^*$. Since $M^*\succ 0$ and all values in $\C_{k-1}$ are non-negative with $0\ \in \C_{k-1}$ being simple, (L1) and (L2) hold for $H(M^*)$. The condition of (L3) is ensured by constraint (\ref{eqn:Phi}). Since $\L_{k-1}$ satisfies (A1) and $M^* \succ 0$, the second largest eigenvalue of $H(M)$ is also strictly greater than 0 which implies $H(M^*)$ satisfies (A1).

Clearly, minimizing $rank(\lambda I -M)$ in (\ref{eqn:minrank}) is equivalent to maximizing the dimension of the nullspace of $\lambda I -M$, which in turn leads to $\lambda$ having the largest multiplicity. However, rank minimization is a well-known difficult numerical problem. The numerical experiment undertaken (see section \ref{sec:num}) suggests that the nuclear norm approximation appears to be the most reliable since it results in a convex optimization problem and is known to be the tightest pointwise convex lower bound of the rank function (\cite{ART:RFP10}).  If such a relaxation is taken, the optimization problem over $\lambda \in \mathbb{R}$ and $M \in \S^{\bar{q}_{k-1}}_{+}$ becomes
\begin{subequations}\label{eqn:nuclearnorm2}
\begin{align}
OPA(\Phi_{k-1}, b_{k-1}): \; &min_{\lambda, M} \|\lambda I - M\|_* \label{eqn:minnuclear2}\\
&M \succ \epsilon_M I_{\bar{q}_{k-1}} \label{eqn:Mepsilon2}\\
&\Phi_{k-1} vec(M) = b_{k-1} \label{eqn:Phi2}
\end{align}
\end{subequations}
where $\epsilon_M$ is some small positive value to prevent eigenvalues of $M$ being too close to $0$. Suppose ($\lambda^* , M^*$) is the optimizer of OPA. There are many cases where the solution of OPA provides a low value of $rank(\lambda^* I -M^*)$. However, there are also many cases where their solutions differ.  This is not unexpected since the nuclear norm is a relaxation of the rank function. In one of these cases, further progress can be made. This special case is characterized by $M^*$ having several eigenvalues that are relatively close (known hereafter as bunch eigenvalues) to one another but are not close enough for
the nullspace of $(\lambda^* I - M^*)$ to have a dimension greater than one. When this situation is detected, a correction step is invoked. Specifically, suppose $spec(M^*)=\{\mu_1,\cdots, \mu_{\bar{q}_{k-1}}\}$ and there are $\ell$ bunch eigenvalues with $\bar{q}_{k-1} \ge \ell \ge 2$ in the sense that
\begin{align}\label{eqn:lambdastar}
|\lambda^* - \mu_i| < \epsilon_\mu \quad \forall i =1,\cdots, \ell
\end{align}
where $\epsilon_\mu >0$ is some appropriate tolerance, then the following correction step is invoked. For notational simplicity, its description uses $q$ and $\bar{q}$ for $q_{k-1}$ and $\bar{q}_{k-1}$ respectively. The input to the correction step are $\lambda^*, M^*,\ell$ from (\ref{eqn:nuclearnorm2}) and (\ref{eqn:lambdastar}).

Correction step: COS($\lambda^*, M^*,\ell$)
\begin{description}
  \item Step 1: Let $\eta=0$ and $(\lambda^* I - M^*)$ is approximated by a rank $\bar{q}-\ell$ matrix of the form
 $$\lambda^* I - M^* \approx F_\eta G_\eta'$$
via full rank matrices $F_\eta, G_\eta \in \mathbb{R}^{\bar{q} \times (\bar{q}-\ell)}$.
  \item Step 2: Solve the following optimization problem over variables $\lambda \in \mathbb{R}, M \in \S^{\bar{q}}_+, \Delta_G, \Delta_F \in \mathbb{R}^{\bar{q} \times (\bar{q}-\ell)}$
\begin{subequations}\label{eqn:nuclearnorm3}
\begin{align}
OPC(\Phi_{k-1}, b_{k-1}): &min \|\lambda I - M - F_\eta G_\eta' - F_\eta \Delta_G' - \Delta_F G_\eta \|_F \label{eqn:minnuclear3}\\
&M \succ \epsilon_M I_{q_{k-1}} \label{eqn:Mepsilon3}\\
&\|\Delta_G\|_F \le \epsilon_G, \quad  \|\Delta_F\|_F \le \epsilon_F \label{eqn:DeltaG}\\
&\Phi_{k-1} vec(M) = b_{k-1} \label{eqn:Phi3}
\end{align}
\end{subequations}
\indent \indent and let its optimizers be $\lambda_\eta^*, M_\eta^*, \Delta_F^*, \Delta_G^*$.
  \item Step 3:  If $\|\lambda_\eta^* I - M_\eta^* - F_\eta G_\eta' - F_\eta (\Delta_G^*)' - (\Delta_F^*) G_\eta\|_F < \epsilon_C \cdot \bar{q}$ or $\eta \ge \eta_{max}$, then stop.
  \item Step 4:  Else, Let $F_{\eta+1}=F_\eta + \Delta_F^*$, $G_{\eta+1}=G_\eta + \Delta_G^*$, $\eta=\eta+1$ and goto Step 2.
\end{description}
The motivation of the correction step is clear. When $F_\eta$ and  $G_\eta$ are of rank ${\bar{q}-\ell}$, so is $F_\eta G_\eta' + F_\eta \Delta_G' + \Delta_F G_\eta$. Hence, when $\|\lambda_\eta^* I - M_\eta^* - F_\eta G_\eta' - F_\eta (\Delta_G^*)' - (\Delta_F^*) G_\eta\|_F $ is sufficiently small, $\lambda_\eta^* I - M_\eta^*$ has a nullspace of dimension $\ell$. The use of full rank matrices, $F$ and $G$ to find a rank $\ell$ solution have appeared in the literature (\cite{INP:FHB04}). However, its use as a correction step after nuclear norm optimization is novel, to the best of the authors' knowledge.

\begin{remark}
The successful termination of COS depends critically on the choices of $F_0, G_0'$. While some options exist, the choice adopted is to let $F_0=G_0'=U_{\bar{q}-\ell} \Sigma_{\bar{q}-\ell}^{0.5}$ where $U_{\bar{q}-\ell}$ is the first ${\bar{q}-\ell}$ columns of $U$, $\Sigma_{\bar{q}-\ell}^{0.5}=diag\{\sigma_1^{0.5}, \cdots, \sigma_{\bar{q}-\ell}^{0.5}\}$ with
$\sigma_1, \cdots, \sigma_\ell$ being the largest $\ell$ singular values of $(\lambda^* I -M^*)$ and $U$ being the corresponding singular vectors.
\end{remark}

As mentioned earlier, the other subalgorithm at iteration $k$ is OPB. OPB is similar to OPA except that $\lambda$ is not a variable. Instead, $\lambda$ is a prescribed value taken successively from $\C_{k-1}\backslash\{0\}$. The need for such a step arises from the nonlinear nature of the rank function. Numerical experiment suggests that the same eigenvalue can be obtained from $M$ eventhough this eigenvalue has been obtained from OPA or COS in an earlier iteration. Hence, the intention of OPB is to check if additional multiplicities can be added to  those eigenvalues in $\C_{k-1}\backslash \{0\}$. Specifically, the optimization problem is
\begin{subequations}\label{eqn:nuclearnorm4}
\begin{align}
OPB(\lambda,\Phi_{k-1},b_{k-1}): \quad \quad min_{M} \; &\|\lambda I - M\|_* \label{eqn:minnuclear4}\\
&M \succ \epsilon_M I_n \label{eqn:Mepsilon4}\\
&\Phi_{k-1} vec(M) = b_{k-1} \label{eqn:Phi4}
\end{align}
\end{subequations}

\begin{corollary} \label{lem:prop2}
Suppose $\L_{k-1}$ satisfies (L1)-(L3) and (A1). Then (i) OPA$(\Phi_{k-1},b_{k-1})$ and OPB$(\lambda,\Phi_{k-1},b_{k-1})$ have a feasible solution, (ii) $H(M^*)$ where $M^*$ is the optimizer of OPA$(\Phi_{k-1},b_{k-1})$ or OPB$(\lambda,\Phi_{k-1},b_{k-1})$ satisfies (L1)-(L3) and (A1). (iii) If OPC$(\Phi_{k-1},b_{k-1})$ terminates successfully at step 3 of COS with $\lambda_\eta^*, M_\eta^*, \Delta_F^*, \Delta_G^*$, then $H(M_\eta^*)$ satisfies (L1)-(L3) and (A1).
\end{corollary}
\textit{Proof} The proof follows same reasoning as those given in the proof of Lemma $\ref{lem:prop1}$.

\section{The Overall Algorithm} \label{sec:algo}
The main algorithm can now be stated.

\begin{algorithm}[H]
\caption{The Minimal Polynomial Algorithm}
\textbf{Input:} $\A_s(G)$ (the standard adjacency matrix of graph G), $\epsilon_M, \epsilon_G, \epsilon_F, \epsilon_C, \epsilon_\mu.$\\
\textbf{Output:} $\L(W_k,G)$, $\C_k=\{\lambda_1,\cdots,\lambda_k\}$ and $\M_k=\{m_1, \cdots, m_k\}$. \\
\textbf{Initialization} \\
Extract $V(G), \E(G)$ from $\A_s(G)$.
Let $\L_0=\L_s$, the standard Laplacian matrix of (\ref{eqn:Lii}) for graph $G$.\\
Set $\C_0=\{0\}, \M_0=\{1\}$, $\xi(0)=1$ and $k=1$. \\
\textbf{Main}
\begin{description}
  \item[1] Compute the eigen-decomposition of $\L_{k-1}=Q_{k-1} \Lambda_{k-1} Q_{k-1}'$ according to (\ref{eqn:Lkminus1}). Set up $\Phi_{k-1}, b_{k-1}$ according to (\ref{eqn:HM}). Compute $q_{k-1}, \bar{q}_{k-1}$ according to (\ref{eqn:qkminus1}). Call OPA$(\Phi_{k-1}, b_{k-1})$ and denote its optimizer as $(\lambda_A^\dag,M_A^\dag)$ with $spec(M_A^\dag)=\{\mu_1,\cdots, \mu_{\bar{q}_{k-1}}\}$. Let $r_A=\bar{q}_{k-1}$.
  \item[2] If (\ref{eqn:lambdastar}) is satisfied with $\ell \ge 2$, then Call COS$(\lambda_A^\dag,M_A^\dag, \ell)$. If COS terminates successfully, let the optimizer be $(\lambda_A^*,M_A^*)$ and $r_A = \bar{q}_{k-1}-\ell+1$.
  \item[3] If $\xi(k-1)=1$, let $r_B=\bar{q}_{k-1}$ and goto step 5. Else, let $n_\C = \xi(k-1)$.
  \item[4] For each $i=2, \cdots, n_\C$,
\begin{description}
  \item[(i)] call OPB($\lambda_i,\Phi_{k-1}, b_{k-1})$ and denote its optimizer as $M_i^\dag$ with $spec(M_i^\dag)=\{\mu_1,\cdots, \mu_{\bar{q}_{k-1}}\}$. Let $r^i_B = \bar{q}_{k-1}$.
  \item[(ii)] If (\ref{eqn:lambdastar}) is satisfied with $\ell \ge 1$, then call COS$(\lambda_i, M_i^\dag, \ell)$. If COS terminates successfully, let its optimizer be $M_i^*$ and $r^i_B = \bar{q}_{k-1}-\ell$.
\end{description}
Next $i$\\
Let $i_B^*=\arg \min_{i=2,\cdots, n_\C}r^i_B$ and $r_B = r_B^{i_B^*}$. If $r_B \le \bar{q}_{k-1}-1$, let $(\lambda_B,M_B)=(\lambda^{i_B^*},M_{i_B^*}^*)$.
  \item[5] If $r_A = \bar{q}_{k-1}$ and $r_B = \bar{q}_{k-1}$, then the algorithm terminates.
  \item[6] If $r_A < r_B$, then let $(\lambda^*,M^*)= (\lambda_A^*,M_A^*)$,
$\C_{k}=\C_{k-1} \cup \{\lambda^*\}$, $\xi(k)=\xi(k-1)+1$, $m_{\xi(k)}= \bar{q}_{k-1}-r_A+1$,  $\M_{k}=\M_{k-1}\cup \{m_{\xi(k)}\}$. Else, let $(\lambda^*,M^*)= (\lambda_B^*,M_B^*)$, $\C_{k}=\C_{k-1}$, $\xi(k)=\xi(k-1)$, $m_{i_B^*}= m_{i_B^*} + (\bar{q}_{k-1}-r_B)$, $\M_k=\M_{k-1}$.
 \item[7] Let $\L_k=H(M^*)$ where $H(\cdot)$ is that given by (\ref{eqn:HM}) and $k=k+1$. Go to 1.
\end{description}

\end{algorithm}

In the above description, quantities $r_A, r_B^i, r_B$ are the estimates of the rank of $M_A, M_B^i$ and $M_B$ respectively. Note that the ranks of $M_A, M_B^i$ and $M_B$ are never computed exactly. Their values are only guaranteed via the successful termination of the COS routine, as given by steps 2 and 4(ii) respectively.

\section{Numerical Examples}\label{sec:num}
This section begins with the examples of graph with known minimal polynomials. In particular, a complete graph, a star-shaped graph and a special regular graph (\cite{ART:VH98}) are used. For each, the minimal polynomials based on the standard Laplacian are used as a reference for the output of Algorithm 1. The parameters in Algorithm 1 are: $\epsilon_M=0.01, \epsilon_G=0.01, \epsilon_F=0.01, \epsilon_C=10^{-7}, \epsilon_\mu=0.01.$  Most of the computations in this section can be done in tens of seconds except when $n=50$ where the computational times are in the range of 5-10 minutes on a Windows 7 PC with a Intel Core i5-3570 processor and 8GB memory. The matlab implementation of this code is available in \cite{MISC:WO17}.

\begin{figure}[H]
\centering
\subfigure[The complete graph]{\label{fig:complete}\includegraphics[width=1.6in]{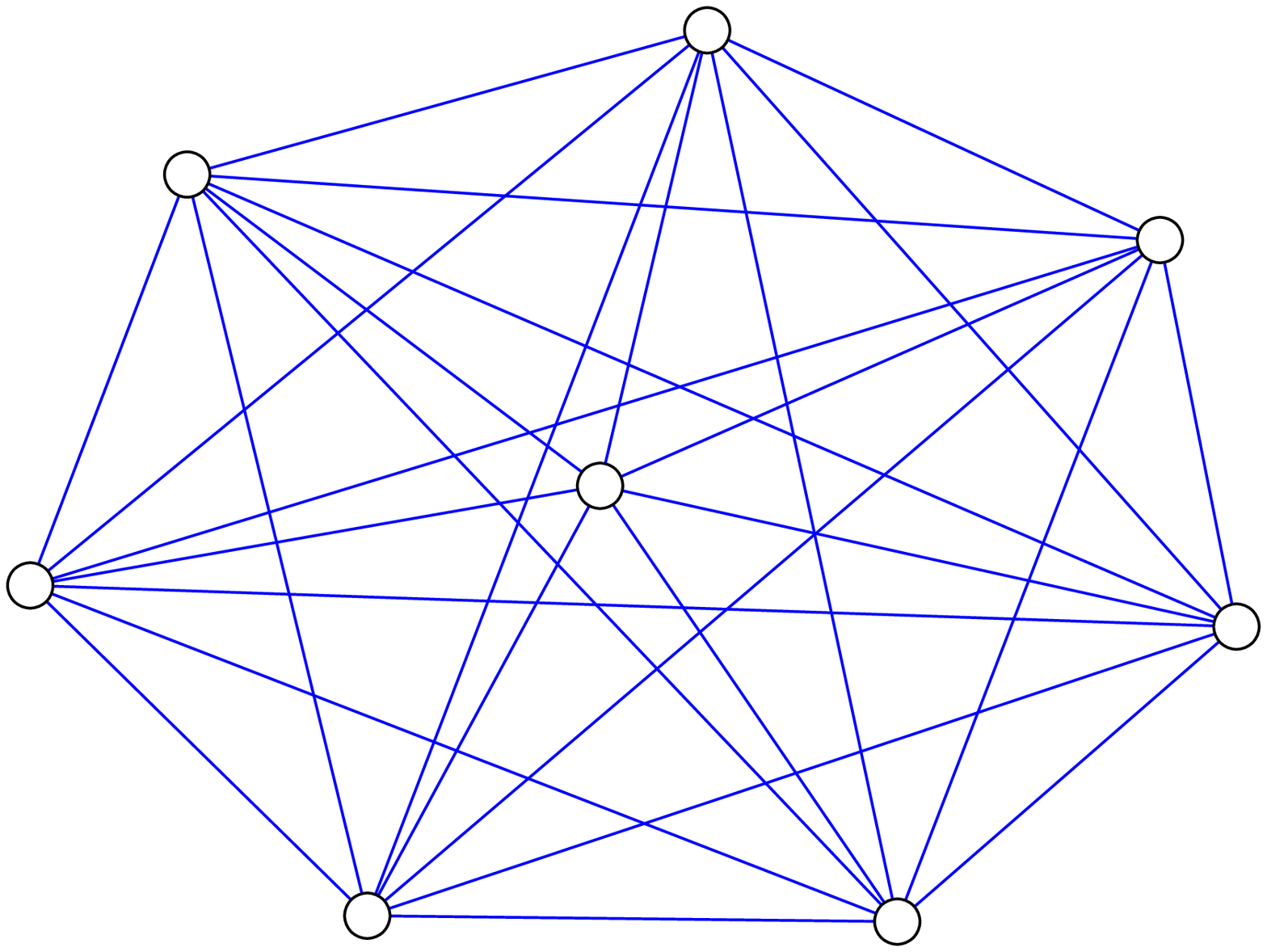}}
\subfigure[The star]{\label{fig:star}\includegraphics[width=1.6in]{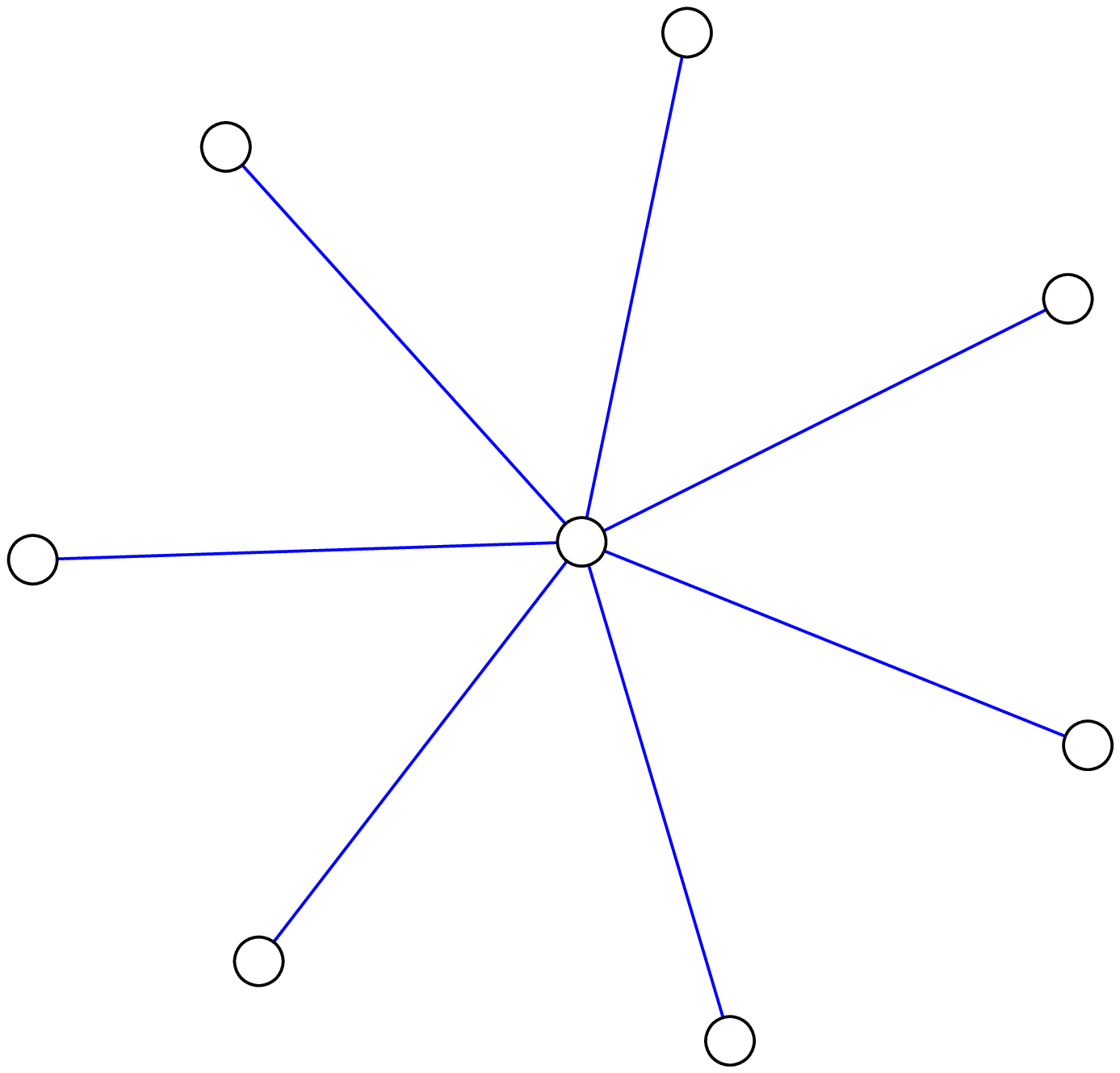}}
\subfigure[The regular graph]{\label{fig:regular}\includegraphics[width=1.6in]{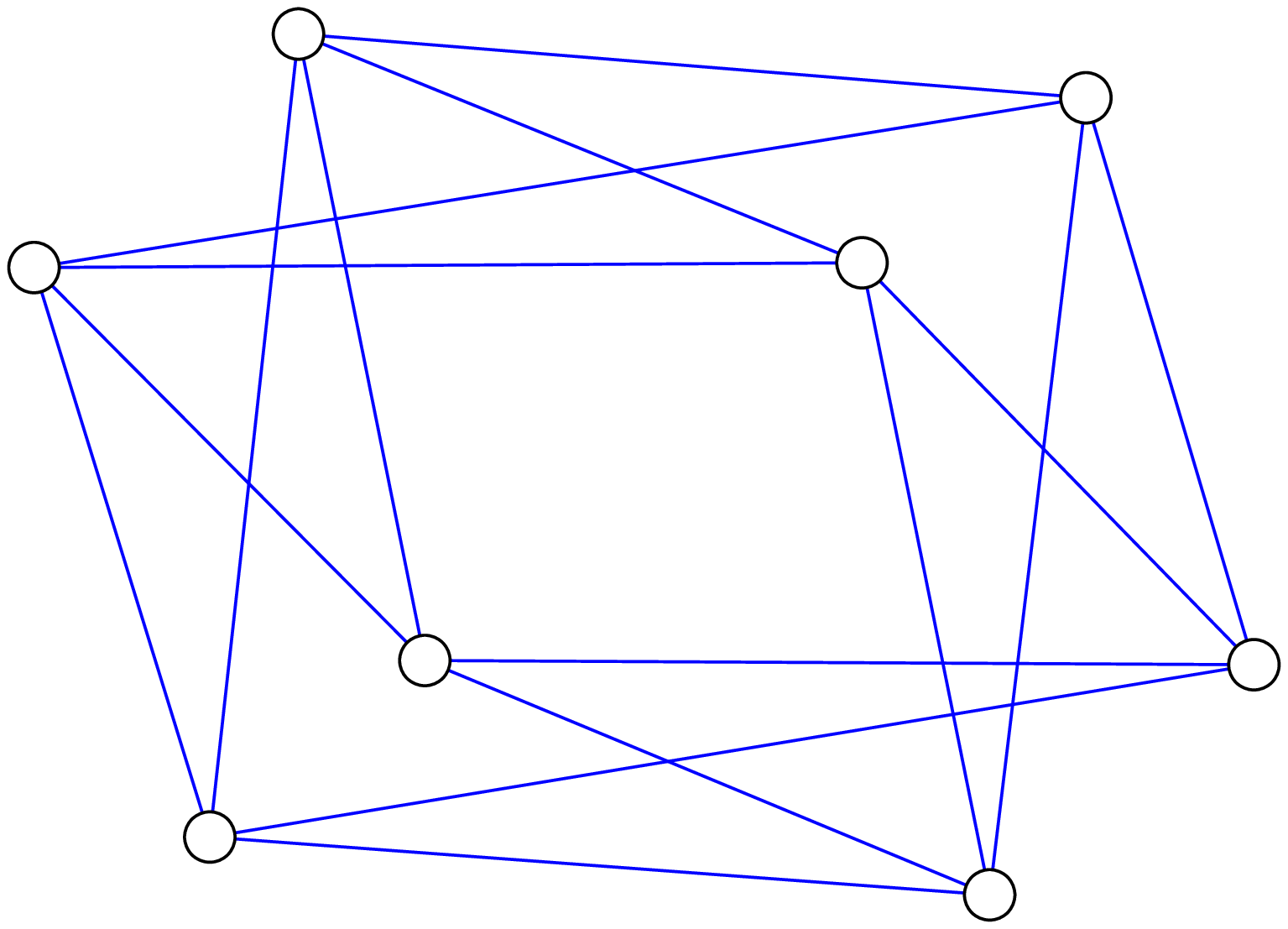}}
\caption{Special examples}
\label{fig:regularstar}
\end{figure}

The spectra of the standard Laplacian matrices of these graphs are: $\{0,8,8,8,8,8,8,8\},\{0,1,1,1,1,1,1,8\}$, and $\{0,4,4,4,4,4,4,8\}$ respectively. The corresponding spectra of $\L(W,G)$ after Algorithm 1 are: \\ $\{0,0.2610,0.2610,0.2610,0.2610,0.2610,0.2610,0.2610\}$, $\{0,7.9982,0.9998,0.9998,0.9998,0.9998,0.9998,0.9998\}$\\
and $\{0,2.0006,1.0003,1.0003,1.0003,1.0003,1.0003,1.0003\}$ respectively. Hence, the algorithm preserves the order of the minimal polynomials for these cases.

\begin{figure}[H]
  \centering
  \includegraphics[width=0.5\linewidth]{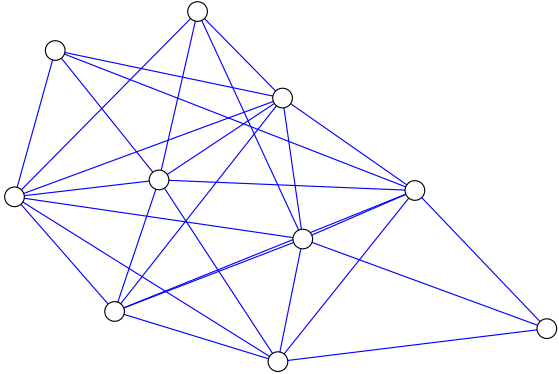}
\caption{A random 10-agent network}
\label{fig:10agent}
\end{figure}

The next example is a 10 agent system with a randomly generated topology as given in Figure \ref{fig:10agent}. It is used to illustrate the progress of a $\L(W,G)$ of a typical graph as it goes through Algorithm 1 as well as a means to evaluate the relevance of the subroutines involved. The next table shows such a case. The second column (labeled Defining Step) of Table \ref{tab:1} refers to the procedure that determines the $M$ matrix of $H(M)$. As the table shows, all routines (OPA, COS-of-OPA, OPB and COS-of-OPB) are needed to achieve the  minimal polynomial. It also validates the necessity of OPB and COS-of-OPB in the algorithm.

\begin{table}[H]
\begin{center}
\begin{tabular}{|c|c|c|c|}
  \hline
  k & Defining Step & Spectra of $H(M^*)$ & order of $m_{\L_k}$ \\
  \hline
  0 & Standard Laplacian & \{0,2.5721,3.7509,4.5858,6.6243,7.1464,7.4142,8,8.8035,9.1028\} & 10 \\
  \hline
  1 & COS of OPA & \{0,2.7246,1.0923,0.9992,0.9995,2.2654,2.1183,2.0839,2.0839,2.0839\} & 8 \\
  \hline
  2 & COS of OPB & \{0,2.9853,0.7331,1.3771,1.5191,2.3286,2.0839,2.0839,2.0839,2.0839\} & 7 \\
  \hline
  3 & COS of OPA & \{0,0.7464,2.9379,2.2874,1.3771,1.3771,2.0839,2.0839,2.0839,2.0839\} & 6 \\
  \hline
  4 & Terminated & \{0,0.7464,2.9379,2.2874,1.3771,1.3771,2.0839,2.0839,2.0839,2.0839\} & 6\\
  \hline
\end{tabular}
\caption{The Steps in Algorithm 1 for the graph of \ref{fig:10agent}} \label{tab:1}
\end{center}
\end{table}

The next table shows results from Algorithm 1 on graphs for which the minimal polynomials are unknown. These graphs are generated based on the following procedure. For every pair of nodes in $\V$, the existence of a link connecting them follows a uniform density function with a threshold. The link exists if and only if the density function returns a value above the threshold.  Graphs of various sizes and topologies are generated in this way. For each graph $G$ generated, validity of assumption (A1) is ensured by checking that the second smallest eigenvalue of $\L_s$ satisfy $\lambda_2 (\L_s(G)) >0$. For each choice of threshold and size, $20$ examples are generated randomly. Let $s_{\L(W)}$ denote the order of $m_{\L(W)}$ from Algorithm 1 and recall that $s_{\L(W)}-1$ is the number of steps needed to achieve consensus from (\ref{eqn:sumtoT}). The mean and standard deviations of $s_{\L(W)}-1$ over the 20 examples are given in the following table:
\begin{table}[H]
\begin{center}
\begin{tabular}{|c|c|c|c|c|c|}
  \hline
  Threshold & $n$ & mean of $s_{\L(W)}$ & std. dev. of $s_{\L(W)}$ & mean of $s_{L_s}$ & std. dev. of $s_{L_s}$ \\
  \hline
  0.3 & 10 & 5.45  & 0.865 & 7.55 & 1.43\\
  0.6 & 10 & 8.5  & 0.5 & 9.95 & 0.218\\
  \hline
  0.3 & 20 & 7.85 & 1.96 & 19.8 & 0.536\\
  0.6 & 20 & 16.9 & 0.624 & 20 & 0\\
  \hline
  0.3 & 50 & 19.3 & 1.58 & 50 & 0\\
  0.6 & 50 & 40.1 & 1.26 & 50 & 0\\
  \hline
\end{tabular}
\caption{The mean and standard deviation of the order of the minimal polynomial over 20 random graphs} \label{tab:2}
\end{center}
\end{table}

Two trends are clear from Table \ref{tab:2} besides the obvious that the order of the minimal polynomial increases for increasingly sparse networks. Let $\bar{s}_{{\L(W)}}$ denote the mean of $s_{\L(W)}$ (third column of Table \ref{tab:2}). The first is that the relative decrease in  $\frac{\bar{s}_{\L(W)}}{n}$ from dense to sparse networks increases for increasing values of $n$. It went from $0.85$ to $0.54$ when $n=10$ and $0.8$ to $0.386$ when $n=50$. This suggests that the proposed approach is more effective for larger networks. The second trend is that percentage of decrease in the order of the minimal polynomial is more pronounced for better-connected system - the ratio of $\frac{\bar{s}_{\L(W)}}{n}$ increases from 0.545 to about 0.386 when $n$ increases from 10 to 50. The percentage decrease is less in the case for sparsely-connected networks, $\frac{\bar{s}_{\L(W)}}{n}$ goes from 0.85 to 0.8 for the corresponding increase in $n$. This suggests that the difficulties in reducing the order of minimal polynomial for sparesely-connected networks.

\section{Conclusions}\label{sec:con}
This work presents an approach to speed up finite-time consensus by searching over the weights of a weighted Laplacian matrix.
The intention is to find the weights that minimizes the rank of the Laplacian matrix. As rank minimization is a difficult problem, this work
uses an iterative process wherein two optimization problems are solved at each iteration. In each optimization problem, a nuclear norm convex optimization is first solved followed by a correction step via a low-rank approximation. Numerical experiment suggests that this two-stage process is more effective in finding a low rank Laplacian matrix.

The numerical experiment shows that the minimal polynomials obtained from the iterative process are of a lower or equal order to that obtained from standard Laplacian. Using the minimal polynomial so obtained, finite-time consensus can be achieved in $(s_\L -1)$ time step where $s_\L$ is the order of the minimal polynomial.


\bibliographystyle{dcu}
\bibliography{Reference}







\end{document}